\documentclass[lettersize,journal]{IEEEtran}
\usepackage{amsmath,amsfonts}
\usepackage[linesnumbered,ruled,vlined]{algorithm2e}
\usepackage{array}
\usepackage[caption=false,font=normalsize,labelfont=sf,textfont=sf]{subfig}
\usepackage{textcomp}
\usepackage{stfloats}
\usepackage{url}
\usepackage{verbatim}
\usepackage{graphicx}
\usepackage{cite}
\usepackage{xcolor}

\newcommand{\eur}{\texteuro\,}

\begin{document}

\title{Sustainable Wireless Services with UAV Swarms Tailored to Renewable Energy Sources}
\author{Igor Donevski,~\IEEEmembership{Student Member,~ IEEE,}\, 
        Marco Virgili,~\IEEEmembership{Student Member,~ IEEE,}\,
        Nithin Babu,~\IEEEmembership{Student Member, ~IEEE,}\,
        Jimmy Jessen Nielsen,~\IEEEmembership{~Member,~IEEE,}\,
        Andrew J. Forsyth,\,~\IEEEmembership{Senior Member, ~IEEE,}\,\\
        Constantinos B. Papadias,\,~\IEEEmembership{Fellow, ~IEEE,}\,
        Petar Popovski,\,\IEEEmembership{~Fellow, ~IEEE} 
\thanks{The work was supported by the European Union's research and innovation programme under the Marie Sklodowska-Curie grant agreement No. 812991 ''PAINLESS'' within the Horizon 2020 Program.} 
\thanks{I. Donevski, N. Babu, J. J. Nielsen, and P. Popovski are with Department of Electronic Systems, Aalborg University, Denmark (e-mail:\{igordonevski, niba, jjn, petarp\}@es.aau.dk).}
\thanks{M. Virgili and A. J. Forsyth are with The University of Manchester, United Kingdom (e-mail: marco.virgili@postgrad.manchester.ac.uk, andrew.forsyth@manchester.ac.uk).}
\thanks{M. Virgili is with Lyra Electronics Ltd., United Kingdom (e-mail: mvirgili@lyraelectronics.com).}
\thanks{N. Babu, and C. B. Papadias is with SWIFT lab, Research, Technology and Innovation Network (RTIN), ALBA, 
The American College of Greece, Greece (e-mails: nbabu@acg.edu, cpapadias@acg.edu).}
\thanks{This work has been submitted to IEEE for possible publication. Copyright may be transferred without notice, after which this version may no longer be accessible.}
}

\maketitle

\begin{abstract}
Unmanned Aerial Vehicle (UAV) swarms are often required in off-grid scenarios, such as disaster-struck, war-torn or rural areas, where the UAVs have no access to the power grid and instead rely on renewable energy. Considering a main battery fed from two renewable sources, wind and solar, we scale such a system based on the financial budget, environmental characteristics, and seasonal variations. Interestingly, the source of energy is correlated with the energy expenditure of the UAVs, since strong winds cause UAV hovering to become increasingly energy-hungry. The aim is to maximize the cost efficiency of coverage at a particular location, which is a combinatorial optimization problem for dimensioning of the multivariate energy generation system under non-convex criteria. We have devised a customized algorithm by lowering the processing complexity and reducing the solution space through sampling. Evaluation is done with condensed real-world data on wind, solar energy, and traffic load per unit area, driven by vendor-provided prices. The implementation was tested in four locations, with varying wind or solar intensity. The best results were achieved in locations with mild wind presence and strong solar irradiation, while locations with strong winds and low solar intensity require higher Capital Expenditure (CAPEX) allocation.

\end{abstract}


\section{Introduction}
\label{introduction}

{The use of UAVs, in particular the multi-copter drones, has been praised for the ability of providing modular, adaptable and scalable wireless communications services as they can easily be redeployed, target specific users and load balance existing cellular architectures, \cite{tutorial,mainsurvey}. Unfortunately, UAV-mounted small base stations (UAVSBSs) are not a feasible replacement to traditional base stations in urban areas, mainly due to safety, privacy and noise concerns. Opposed to this, UAVSBSs are crucial in scenarios that result in service outages such as war-torn or disaster-struck areas \cite{disaster} and traffic surges in weakly serviced areas \cite{multitier}. In these cases, it should be expected that the existing infrastructure is unable to support the energy requirements of the UAVSBS system. Moreover, to satisfy the service constraints of the area, which generally vary during the day \cite{donevski2019neural}, the deployments require multiple UAVs (a.k.a. swarm). 

Since UAVs require a lot of energy (stored in a battery) to fly, the goal of this work is to evaluate the feasibility of self-sustainable energy systems for long-term persistent (uninterrupted) operation of UAV swarms. This is targeted for areas where fixed infrastructure is unavailable and a UAVBS-based solution is deemed acceptable. As such, we aim to provide a design solution by finding the ideal scale of the system for a particular location, while maximizing the coverage area discounted by its financial cost. This is a nuanced problem, as it solves complex interactions between the energy generation and consumption systems.}

\subsection{Literature Overview}
The effect of using UAVSBSs that are positioned to offer service to ground customers has already been well investigated in \cite{mainsurvey,dynamic,DCs,foundations,igoreucnc,donevski2021standalone,3dplace,atg,optlap}. In \cite{dynamic,DCs,foundations} the focus  is on improving spectral efficiency when exploiting the temporal and spatial mobility of UAVs for servicing user hotspots. In our previous works \cite{igoreucnc,donevski2021standalone}, we demonstrated the benefit of horizontally positioning a standalone UAVSBS, equipped with a tilting directional antenna. Moreover, the work in \cite{3dplace} focused on the energy efficiency for UAVSBS deployment, while the authors in \cite{babu1} and \cite{babu2} studied the problem of placement optimization of a single cell and interference-limited multi UAVSBS deployments, respectively. While the aforementioned works are concerned with optimizing deployment locations of the UAVs once they are in the air, they generally ignore the problem of short service time.

{
Ever since the proliferation of drones into the mass market, there has been a strive towards persistent UAV services \cite{7743546} with several methods. The most prominent one assumes automated battery swapping \cite{michini2011automated}. In \cite{batswaptraj}, the authors solve the optimal trajectory for patrolling UAVs that thoroughly exploits the battery swapping mechanism connected to the grid mains. In \cite{sillymothership}, the authors consider a mother ship-like UAV that houses and orchestrates the deployment of a swarm of smaller UAVs, where the mother ship  ensures that the energy requirements for the entire system are satisfied. While such mother ship systems are genuinely useful for achieving unlimited mobility, the creation of one is complex and it assumes technical innovation on several fronts, which is a significant shortcoming and would become very costly to implement. On the other hand, the authors of \cite{GALANJIMENEZ2021102517} consider a ground-based central unit that serves as a backbone to the UAVs and has solar panels to manage the energy requirements on the ground. The shortcoming of the previous work is that it does not consider the impact of wind and the energy needed to offset it. In \cite{coverageopt}, the authors propose a cost-efficient UAV system for data harvesting from IoT systems, but this is not directly related to general communication services. In \cite{nithinTVT}, we previously investigated the optimal arrangement for UAVs that need to provide persistent service by interleaved recharging at a ground station. However, the analysis was limited to a single UAV, and without the impact of wind.
\begin{figure*}[t]
\setlength{\belowcaptionskip}{-5pt}
\centering
\centerline{\includegraphics[width=0.96\linewidth]{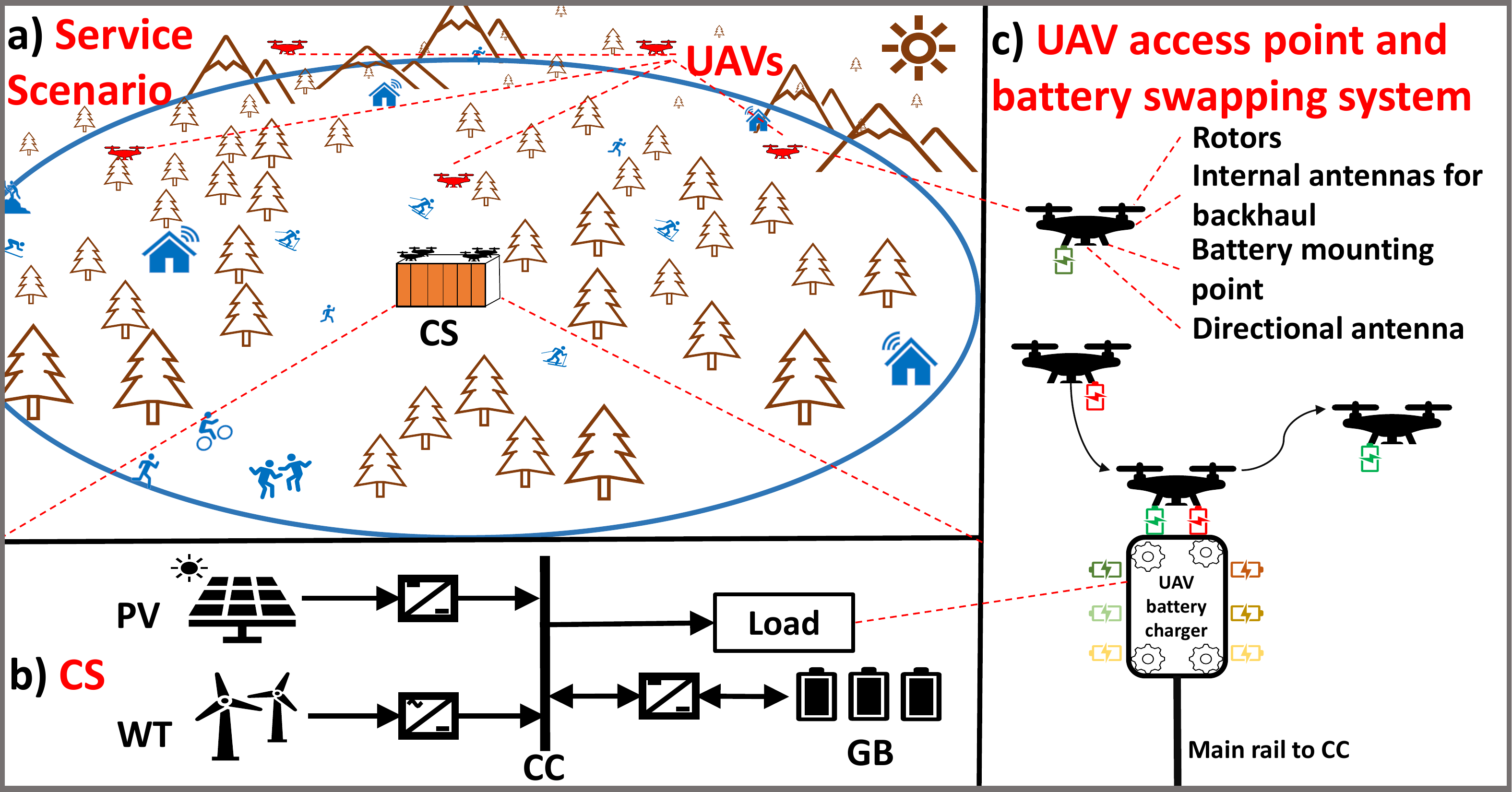}}
\caption{Schematic of the energy system at the central station (CS) that consists of wind turbines (WT), photovoltaic (PV) panels, a central circuit (CC), ground battery (GB) and a UAV battery charger (UAVBC) that represents the load. The service scenario illustrates a deployment in a mountainous region.}
\label{fig:ensys}
\end{figure*}

We note that per \cite{mainsurvey}, UAV Base Stations are able to alleviate capital and operating expenditures (CAPEX \& OPEX) of telecom operators up to 52\% and 42\%, respectively. 
To serve disaster-struck, remote, or underdeveloped areas, we focus on works that involve sizing sustainable energy generation systems for wireless communications. As such, the authors of \cite{sushetnets} and \cite{renewablemarsan} proposed alleviating the energy requirements of multi-tier cellular implementations supported with renewable energy. The work of \cite{mincostdesign} comes the closest to our goal of providing cellular connectivity in rural zones. Moreover, the authors consider an architecture composed of UAV-based BSs to provide cellular coverage, ground sites to connect the UAVs with the rest of the network, solar panels, and batteries to recharge the UAVs. The \cite{mincostdesign} approach is generally simplistic, does not maximize coverage, and does not account for the impact of wind. And, finally, the work \cite{amorosi2021coordinating} analyzes a mother ship-orchestrated UAV swarm for wireless communications, where the goal is to minimize the overall weighted distance traveled by the mother ship for UAV recharge. This work does not provide a realistic overview of the capital expenses, and omits the impact of wind.}

\subsection{Off-grid Redeployable UAV Communications System}
The proposed communications system is shown in Fig.~\ref{fig:ensys}, and is intended to provide persistent services to rural, suburban and low-rise urban areas, by deploying a central station (CS) that supports and coordinates the UAV swarm. Once deployed, the UAVs hover and provide a satisfactory service rate for the entire area. When a UAV nearly depletes its battery, it flies back to the CS to replace the  energy spent. As in \cite{batteryswap}, we consider an automatic battery swapping mechanism that replaces the depleted batteries, as shown in c) in Fig.~\ref{fig:ensys}.

To ensure a long-term persistent deployment, the system needs to compensate the power requirements. As illustrated in Fig.~\ref{fig:ensys}, our proposed implementation of a CS has five energy modules. The wind turbines (WT) and photovoltaic (PV) panels generate energy to be stored into a central ground battery (GB). The system is interconnected by a central circuit (CC) module that directly links the load of the system, which is an automatic UAV battery charger (UAVBC) that charges the hot-swappable UAV batteries. Finally, we assume that the CS acts as a sink/middle-haul for the wireless service that the UAVs offer, which is over-provisioned and provided by low earth orbit (LEO) satellites \cite{satbackhaul}. 

\subsection{Main Contributions \& Paper Outline}

{This work provides a fresh perspective on UAV swarm implementations for persistent wireless service, such as: 
\begin{itemize}
    \item We consider long-term standalone deployment of UAVs for remote areas with a realistic model for the impact of wind onto the energy consumption of the UAVs. This is the first work in the area to consider wind intensity, accounting for both altitude and terrain roughness.
    \item We model the system as totally self-sustainable and including PV- and WT-based energy generation modules. This is the first work to introduce WTs that have the capability of offsetting the UAV's extra energy expenditure due to wind speed.
    \item We aim to provide coverage in the most efficient manner for a potential capital investment. To achieve this, we formulate a novel problem that maximizes the wireless coverage area discounted by the cost of the system. The goal of solving the problem as coverage maximization is to understand the scalability of such a complex system in the many different possible deployment environments. This is a nuanced multi-variate optimization problem and it is entirely based on real world data and current commercial climate.
    \item We propose a computationally light algorithm that uses greedy sampling and binary search to find the optimal configuration that is a combination of covered area, wind turbines, PV panels, cells in the ground battery, and UAVs in the swarm.
\end{itemize}}
Common abbreviations are contained in Tab.~\ref{tab:abbrevs}, and commonly used symbols are contained in Tab.~\ref{tab:nomen}. Symbols introduced later in the papers are contained in a table within that section.
The rest of the paper is organized as follows. In Section~\ref{model}, we describe the UAV based communications services and its energy expenditure. In Section~\ref{enman},
we introduce the energy generation and management system. In Section~\ref{problem}, we define both the formal problem and the proposed algorithmic solution. In Section~\ref{numerical}, we display the results of the implementation, and finally draw the conclusions in Section~\ref{conclusion}.

\begin{table}[]
\caption{Abbreviations used in this work.}
\centering
\begin{tabular}{ll}
\hline
Abbreviation & Meaning \\ \hline \hline
BS & Base Station \\
CA & Coverage Area \\
CC & Central Circuit \\
CCEE & Cheapest Combination of Energy Elements \\
CS & Central Station \\
GSS & Greedy and Sparse Search \\
IoT & Internet-of-Things \\
LoS & Line-of-Sight \\
MEL & Minimum Energy Load \\
MPPT & Maximum Power Point Tracker \\
NLOS & No-Line-of-Sight \\
NOC & Nominal Operating Conditions \\
PV & Photovoltaic \\
QoS & Quality of Service \\
ST & Standard Testing conditions \\
UAV & Unmanned Aerial Vehicle \\
UAVBC & UAV Battery Charger \\
UAVSBS & UAV-mounted Small Base Station \\
WT & Wind Turbine \\
ZDD & Zonal Datarate Density \\
CAPEX & Capital Expenditure \\
\hline
\end{tabular}
\label{tab:abbrevs}
\end{table}

\begin{table}[]
\caption{Nomenclature of the symbols used in this work.}
\centering
\begin{tabular}{ll}
\hline
Symbol & Meaning and unit of measurement \\ \hline \hline
$A_\text{eff}$ & Antennas' effectiveness in fitting the CA \\
$C_\text{bat}$ & Battery capacity (Wh) \\
$D_{j} $ & Horizontal Distance of the $j$-th UAV from CS (m)\\
$D_\text{max}$ & Diameter of coverage area (m) \\
$E_{\text{UAVs},h,i}$ & UAV swarm energy consumption at time $h$ and day $i$ (Wh) \\
EEAC & Energy Efficiency of Annual Coverage ($\text{m}^2$/Wh) \\
$F$ & Total cost of the system (\texteuro) \\
$F_{a}$ & Cost of each type of wind turbine (\texteuro) \\
$F_\text{E}$ & Cost of battery system (\texteuro) \\
$F_\text{PV}$ & Cost of photovoltaic system (\texteuro) \\
$F_\text{UAV}$ & Cost of UAVs (\texteuro) \\
$F_\text{W}$ & Cost of wind system (\texteuro) \\
$h$ & Hour of the day \\
$i$ & Day of the year \\
$k_h$ & Number of UAVs in a swarm \\
$\ell$ & Path loss (dB) \\
$n_a$ & Number of wind turbines of each type \\
$n_\text{PV}$ & Number of solar panels \\
$n_\text{UAV}$ & Fleet size - Number of available UAVs \\
$\mathbf{p}_j$ & Hovering location of the $j$-th UAV \\
$R$ & Instantaneous Data rate (Mbps) \\
$R_{h,\text{min}}$ & Minimum datarate requirement (Mbps) \\
$v_{h,i}^\text{wind}$ & Wind speed at time $h$ and day $i$ (m/s) \\
$\eta$ & Mean large scale fading coefficient (dB) \\
$\theta$ & Elevation angle at the cell edge (\textdegree) \\
$\lambda_h$ & Zonal datarate density (Mbps/$\text{m}^2$) \\
$\tau_\text{fly}$ & Air-time of a UAV (\% of hour) \\
\hline
\end{tabular}
\label{tab:nomen}
\end{table}

\section{Modeling UAV Service and Energy}
\label{model}
The coverage area (CA), which is a circle of radius $D_\text{max}$, contains an arbitrary number of users that we model in terms of zonal datarate density (ZDD). The ZDD is defined as $\lambda_{h}$, which represents the requested datarate per unit of area, in Mbps/m$^2$, for hour of the day $h = 1,2... 24$, which is uniform for the entire area. The goal of ZDD is to properly scale it for larger time-lengths in the order of hours and adapt it per type of residency area, such as in \cite{marsan,donevski2019neural}. This allows scaling the traffic demand for different sizes of $D_\text{max}$ without having to assume a stochastic point process. As a result, the minimum datarate requested for the entire CA $R_{h,\text{min}}$, for hour $h$, is: $
     R_{h,\text{min}}(D_\text{max}) = \lambda_{h} \pi D^2_\text{max}. $
Note that we do not consider different rates between days of the year $i = 1,2, .. 365$, since such a metric is difficult to obtain and challenges the privacy of users. 
Considering a fleet of available drones $n_\text{UAV}$, a swarm size of $k_{h} \leq n_\text{UAV}$ UAVs is released so that each UAV $j$ is given an equal amount of area to serve with rate $R(k_{h},D_\text{max})$. We can thus linearly scale the traffic load on each UAV with the swarm size, so that it satisfies:
 \begin{equation}
\frac{R_{h,\text{min}}(D_\text{max})}{k_h} \leq R(k_{h},D_\text{max}), \,\, \forall h
\end{equation}
under the condition that $\max_h\left(R_{h,\text{min}}\right) \leq n_\text{UAV} \cdot R(n_\text{UAV},D_\text{max})$ is satisfied. The data rate depends on $k_h$ and $D_\text{max}$ because the radius of coverage of each UAV in the swarm varies within the bounds of $0 < D(k_h,D_\text{max}) \leq D_\text{max}$.
\subsection{UAV Hovering Locations}
The coverage region for each UAV in the swarm is a circle of radius $D(k_h,D_\text{max})$, which is derived from a packing algorithm \cite{circle}. In order to avoid leaving any part of the area without service, the circles of individual UAV coverage are packed in an overlapping manner that fully covers the CA. Making each UAV $j \in \{1,2, \, ..\, k_h\}$ equally relevant, we assign the same radius $D(k_h,D_\text{max}) = D_{j} \,\, \forall j$. Thus, as per the packing provided in \cite{circle}, the radius occupies discrete values $D(k_h,D_\text{max}) = \frac{D_\text{max}}{\gamma_{k_h}}$, where $\gamma_{k_h} = 1, 1, 1.1547, \sqrt{2}, 1.641, 1.7988, 2$, for $k_h = 1, 2, 3, 4, 5, 6, 7$, respectively, and $\frac{D_\text{max}}{1+2\text{cos}\left(\frac{2\pi}{k_h-1}\right)}$ for $k_h=8,9,10$.

\begin{figure}[t]
\setlength{\belowcaptionskip}{-5pt}
\centering
\centerline{\includegraphics[width=1.0\columnwidth]{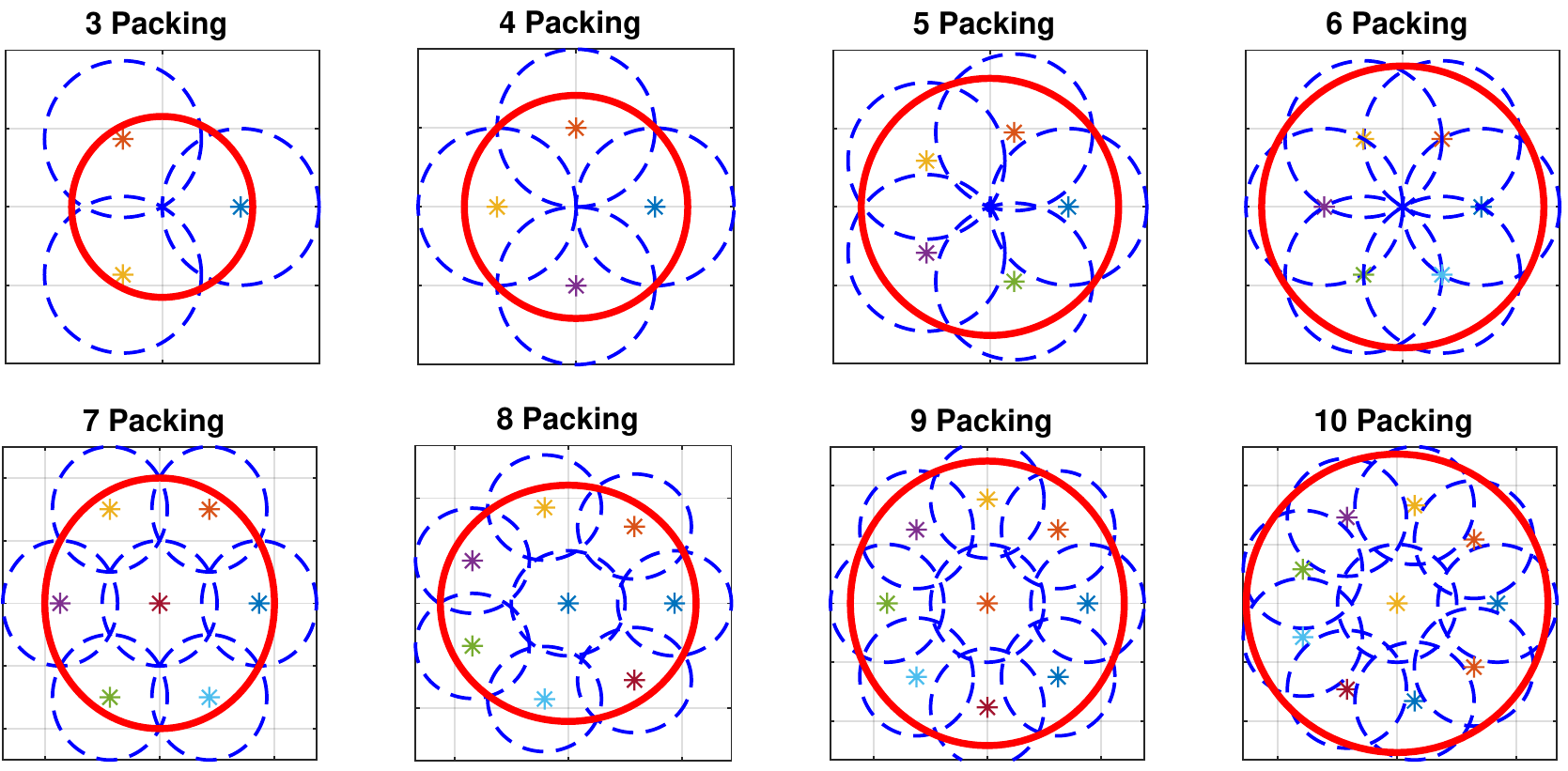}}
\caption{Overlapping packing patterns for UAV regions (blue) with radius $D(k_h,D_\text{max})$ fully covering the CA (red) with radius $D_\text{max}$.}
\label{fig:packing_patterns}
\end{figure}

Setting the center of the CA as the center of our coordinate system (0,0), the centers of the $k_h=\{3,4,5,6\}$ circles are located at $\{\mathbf{p}_j(k_h,D_\text{max})=(x_j,y_j)\}$ where,
\begin{align}
x_{j}=&  D(k_h,D_\text{max})\text{cos}\left(\dfrac{2\pi (j-1)}{k_h-1}\right)\,\, \forall\, j\in \left\lbrace 1,2, ... k_h\right\rbrace,\label{5cirx}\\
y_{j}=&  D(k_h,D_\text{max})\text{sin}\left(\dfrac{2\pi (j-1)}{k_h-1}\right) \,\, \forall\, j\in \left\lbrace1,2, ... k_h \right\rbrace.  \label{5ciry} 
\end{align}\\
For the case of 7, the centers of the smaller circles of radius $D(7,D_\text{max})$ that cover the region have coordinates $\{\mathbf{p}_j(7,D_\text{max})=(x_{j},y_{j})\}$ where, 
\begin{IEEEeqnarray}{lCl}
\small
x_{j}= D(7,D_\text{max})\sqrt{3}\text{cos}(\dfrac{2\pi (j-1)}{6})\,\, \forall\, j\in \left\lbrace1,2,...6\right\rbrace,\label{c1}\\
y_{j}= D(7,D_\text{max})\sqrt{3}\text{sin}(\dfrac{2\pi (j-1)}{6})\,\, \forall\, j\in \left\lbrace1,2,...6\right\rbrace, \\
(x_{7},y_{7})=(0,0).\label{c3}
\end{IEEEeqnarray}
As such, the horizontal distance from the CS can be calculated as $ d_j(k_h,D_\text{max}) = \sqrt{x_j^2+y_j^2}$. Finally, for $k_h=\{8, 9, 10\}$, one circle is concentric with the region and the centers of the other circles are situated in the vertices of a regular $(n-1)$-gon at a distance of $d_j(k_h) = 2\text{sin}(\frac{\pi}{(j-1)}) \,\, \text{for}\,\, j\in \left\lbrace1,2,...(k_h-1)\right\rbrace$ from the center of the region. The circle packing formations are shown in Fig.~\ref{fig:packing_patterns}. In order to achieve coverage regions with radius $D(k_h,D_\text{max})$, we adjust the UAV hovering height $H(k_h,D_\text{max})$, which is dependent on the propagation environment in the CA, and it is covered below.
\subsection{Propagation Characteristics with a Directional Antenna}
UAV based communication links discriminate two propagation groups, users with direct line-of-sight (LoS) or no-LoS (NLoS). As such, the path loss $\ell$ is a sum of the free space path loss (FSPL) and the additional large-scale shadowing coefficient for each one of the propagation groups. The mean large scale fading coefficients for each propagation group are $\eta_\text{LoS}$ and $\eta_\text{NLoS}$ and come as a consequence of the typology's features\cite{atg}. Thus, the path loss between a user at horizontal distance $D$ and a UAV with altitude $H$ can be expressed as:
\begin{align}
\small
 \ell_{\text{\tiny{LoS}}} =& - 10\log(G_\text{t})+ 20\log(\sqrt{D^2 + H^2}) + C + \eta_\text{\tiny{LoS}} ,\\
    \ell_{\text{\tiny{NLoS}}} =& - 10\log(G_\text{t}) + 20\log(\sqrt{D^2 + H^2}) + C + \eta_\text{\tiny{NLoS}},
\end{align}
where $G_\text{t}$ is the antenna gain, $\log$ is a shortened version of the common logarithm $\log_{10}$, and the term $C$ is a substitute for the carrier frequency $f_\text{c}$ constant in FSPL $C =20\log{(\frac{f_\text{c}4\pi}{c})}$. Finally, averaging the two propagation groups by the probability of a LoS occurring gives:
\begin{align}
\label{eq:expected}
    10\log[  L_{}] =  {P}_\text{LoS} (\eta_\text{LoS}-\eta_\text{NLoS}) + \ell_\text{NLoS},
\end{align}
where the LoS probability is given by the s-curve model\cite{optlap}:
\begin{equation}
\label{eq:plos}
    P_\text{LoS} =\frac{1}{1+a\exp(-b[\arctan\left(\frac{H}{D}\right)-a])}, 
\end{equation}
where $a$ and $b$ are constants dependent on the topological setting.

Each UAV has a downwards facing antenna with gain $G_\text{t}=A_\text{eff}10\log(G_\text{I})$, where the ideal conical antenna has gain:
\begin{equation}
\label{eq:dt}
    G_\text{I}  = \frac{2}{1-\sin{(\theta \frac{\pi}{180} )}},
\end{equation}
where $\theta=\arctan(\frac{H}{ D }) $ is the elevation angle at the cell´s edge and $A_\text{eff}$ is the antennas' effectiveness in fitting an ideal conical beamwidth. This results in the final path loss expression:
\begin{multline}
\label{eq:fullos}
 10\log(  L_{} ) = \frac{\eta_\text{LoS}-\eta_\text{NLoS}}{1+a\exp\left\lbrace-b[\theta-a]\right\rbrace} + 20\log\left(\sqrt{D^2 + H^2}\right) \\  - {A_\text{eff}}10\log{\left[\frac{2}{1-\sin{(\theta\frac{\pi}{180})}}\right]} + C + \eta_\text{NLoS}.
\end{multline}
In order for all the users within the area with radius $D$ to be served, we optimize the elevation angle of $\theta=\arctan(\frac{H}{D}) $ from the perspective of a user located exactly at distance $D$. Thus, as in \cite{igoreucnc}, we can extract an optimal ratio of $D$ and $H$, through the angle $\theta$, by solving:
\begin{multline}
\label{eq:dtderiv}
    0 = \frac{\pi \tan{(\theta\frac{\pi}{180}})}{9\log(10)} + \frac{a\,b (\eta_\text{LoS}-\eta_\text{NLoS}) \exp(-b(\theta -a)) }{a\exp(-b(\theta-a)+1)^2 }
    \\- {A_\text{eff}} \frac{\pi  \cos{(\theta\frac{\pi}{180})}}{18\log(10)(1-\sin{(\theta\frac{\pi}{180})} )}.
\end{multline}
This makes it easy to calculate the hovering height as $H= D\tan(\theta)$, which formulates the path loss only as a function of the horizontal distance $L(D)$. Finally, the serving rate for a user at distance $D = D(k_h,D_\text{max})$ becomes:

\begin{equation}
\label{eq:rate}
R(k_h,D_\text{max})  = B\log_2\left[1+\frac{P_\text{t}  }{BN_0 L(D(k_h,D_\text{max}))} \right],
\end{equation}
where $P_\text{t}$ is the transmission power, which is assumed to be identical at both user and UAV side, while $N_0$ is the noise spectral density linearly scaling the noise with the channel bandwidth $B$. Since the packing is done in an overlapping manner, we must account for a total available spectrum of $B_\text{tot} \geq 3 \cdot B$ to avoid inter-UAV-cell interference. Finally, we note that, even though the coverage circles of two UAVs using the same bandwidth may overlap, such overlap occurs outside both coverage regions, and is thus not harmful towards the spectrum reuse in the packing algorithm, as it can be seen in Fig.~\ref{fig:packing_patterns}.

\subsection{UAV Power Consumption Model} 
\begin{table}[]
\caption{UAV Flight Parameters.}
\centering
\begin{tabular}{lll}
\hline
Label & Definition & Value \\ \hline
\hline
$W$ & Weight of the UAV in Newton & 23.84 N \\
$N_{\text{R}}$ & Number of rotors & 4 \\
$F_{\text{n}}$ & Upward thrust by the $n^{\text{th}}$ rotor &  -\\
$v_{\text{hfly}}$ &  UAV's horizontal flying velocity & 10 m/s\\
$v_\text{tip}$ & Tip speed of the rotor & 102 m/s \\
$A_{\text{f}}$ & Fuselage area & 0.038 $\text{m}^2$  \\
$\rho(H(k_h,D_\text{max}))$ & Air density & - \\
$C_\text{D}$ & Drag Co-efficient & 0.9  \\
$A_\text{r}$ & Rotor disc area & 0.06 $\text{m}^2$ \\
$\Delta$ & Profile drag coefficient & 0.002 \\
$s$ & Rotor solidity & 0.05\\
$v_{c}$ &  UAV's vertical flying velocity & 10 m/s\\ 
$P_{\text{vfly}}$ &  Vertical flight power&  -\\
$P_{\text{hfly}}$ &  Horizontal flight power&  -\\
$v_\text{hov}$ & Flying speed to counteract wind (m/s)&  - \\ \hline
\end{tabular}
\label{uavparameters}
\end{table}
Most of the UAV's power consumption is absorbed by its rotors, while the power spent for communications is negligible \cite{nithinTVT}. To hover, the UAV may have to counteract the wind speed $v_{h,i}^\text{wind}$, for hour $h$ at day $i$, to achieve net zero speed is remarked as flying horizontally with non-zero velocity. Here we differentiate the wind intensity with regards to the daily variations, since such data is readily available, and has very strict seasons. We also expect that the horizontal speed required to counteract the wind speed increases with altitude \cite{windpowerlaw}:
\begin{equation}
\small
    v_{\text{hov},h,i}=v_{h,i}^\text{wind} \left[\frac{H(k_h,D_\text{max})}{H_0}\right]^{E_\text{w}},
\end{equation}
where $H_0$ is the measurement altitude of the wind velocity $v_{h,i}^\text{wind}$, and $E_{\text{w}}$ is the empirical coefficient derived relative to the roughness of the surface in the area. 
To reach the hovering position $\mathbf{p}_j(k_h,D_\text{max})$, the UAV ascends vertically with a velocity of $v_{\text{c}}$ to the designated height $H(k_h,D_\text{max})$, and flies horizontally with a velocity of $v_{\text{hfly}}$ the horizontal distance $d_j(k_h,D_\text{max})$. Near the end its air-time $\tau_\text{fly}$, the UAV descends at $-v_{\text{c}}$, that is, with negative velocity with regards to the coordinate system. 

All the parameters used in the following equations are defined in Table~\ref{uavparameters}, and, with the goal to reduce equation clutter, the variables $H(k_h,D_\text{max})$ and $d_j(k_h,D_\text{max})$ are reduced to $H$ and $d_j$, respectively. The power consumed by the UAV when flying horizontally with speed $v$ is derived using the axial momentum theory, while assuming identical rotors \cite{nithinTVT} as,
\begin{IEEEeqnarray}{rCl}\label{phfly}
P_{\text{hfly}}(v)&=&\underbrace{ N_{\mathrm{R}}P_{\mathrm{b}}\left(1+\dfrac{3v_{\text{}}^2}{v_{\text{tip}}^2}\right)}_{P_{\text{blade}}}+\underbrace{\dfrac{1}{2}C_\text{D}A_{\text{f}}\rho(H)v_{\text{}}^3}_{P_{\text{fuselage}}}\nonumber\\
&+&\underbrace{ W\left(\sqrt{\dfrac{W^2}{4 N_{\mathrm{R}}^2 \rho^2 (H)A_{\mathrm{r}}^2}+\dfrac{v_{\text{}}^4}{4}}-\dfrac{v_{\text{}}^2}{2}\right)^{\frac{1}{2}}}_{P_{\text{induce}}},
\end{IEEEeqnarray}
where $P_{\text{b}}=\dfrac{\Delta}{8}\rho(H) s A_{\mathrm{r}} v^3_{\text{tip}}$, $\rho(H)=(1-2.2558.10^{-5} H)^{4.2577}$. $P_{\text{blade}}$ and $P_{\text{fuselage}}$ are the powers required to overcome the profile drag forces of the rotor blades and the fuselage of the aerial vehicle that oppose its forward movement, respectively, while $P_{\text{induce}}$ represents the power required to lift the payload.

The power required by the aerial vehicle to climb vertically with a rate $v_\text{c}$ m/s is expressed as,
\begin{equation}
    \small
     P_{\text{vfly}}(v_\text{c}) = \dfrac{W}{2}\left(v_\mathrm{c}+\sqrt{v_\mathrm{c}^2+\dfrac{2W}{N_{\mathrm{R}}\rho(H)A_\mathrm{r}}}\right)+N_{\mathrm{R}}P_{\mathrm{b}}.
 \end{equation}
 
The energy consumption for the entire flight of UAV $j$ occurring at hour $h$, day $i$, is $E_{j,h,i}(k_h,D_\text{max})$ and can be thus segmented into the three parts, ascent, hovering, and descent:
\begin{IEEEeqnarray}{lCl}\label{eq:energytot}
E_{j,h,i}(k_h,D_\text{max}) = \underbrace{ P_{\text{vfly}}(v_\text{c}) \frac{H}{v_\text{c}}+P_{\text{hfly}}(v_\text{hfly})\frac{d_j}{v_\text{hfly}}}_\text{ascent}\nonumber\\
+ \underbrace{ P_{\text{vfly}}(-v_\text{c}) \frac{H}{v_\text{c}}+P_{\text{hfly}}(v_\text{hfly})\frac{d_j}{v_\text{hfly}}}_\text{descent}\nonumber\\
+ \underbrace{P_{\text{hfly}}(v_{\text{hov},h,i}) \cdot \left(\tau_\text{fly} - 2 \left(\frac{H}{v_\text{c}} + \frac{d_j}{v_\text{hfly}}\right) \right)}_\text{hover},
\end{IEEEeqnarray}
where $\tau_\text{fly}$ is the designated flight time that the UAV must complete, and $2 \left(\frac{H}{v_\text{c}} + \frac{d_j}{v_\text{hfly}}\right) <\tau_\text{fly}$. For convenience, we use a flight duration $\tau_\text{fly}$ of half an hour, which is reasonable  for state-of-the-art UAV models, since our wind, solar and traffic data are quantized at each hour of the day. This means that, at hour $h$ on day $i$, the UAV consumes a total energy of:
\begin{equation}
  E_{\text{UAVs},h,i}(k_{h},D_\text{max})  = \frac{1}{\tau_\text{fly}}\sum_{j=1}^{k_{h}}{E_{j,h,i}(k_{h},D_\text{max})}.
\end{equation}

Finally, we note that some of the flight time is spent on flying to and back from designated hovering positions. 
To avoid service outage and add leeway for battery swapping, we assume that the process of positioning occurs at different times for each UAV. To afford such mobility, the system requires one spare auxiliary UAV.

\section{Energy Generation and Management at the Central Unit}
\label{enman}
The electricity generated and stored in this system is proportional to its size, and therefore to its financial budget. The service availability detailed in the previous section thus becomes a function of the financial budget, which is spent on energy generation and storage systems for the CS.
\subsubsection{Load}
Once the UAV lands on the CS, after spending $\tau_\text{fly}$ time in the air, it releases its depleted battery through an automated battery exchange system and receives a new, fully charged, one, as shown back in Fig.~\ref{fig:ensys} part c). The old battery is then fully charged, making each recharge cycle duration $\tau_\text{charge} =\frac{C_\text{bat}}{P_\text{charge}},$
where ${C_\text{bat}} $ is the battery capacity, and ${P_\text{charge}}$ is the charging power. The lithium polymer (LiPo) on-board batteries have a predominantly linear charging behaviour \cite{alessandrini}. Therefore, the power drawn by a single battery unit is assumed to be constant, and the overall load profile will look like a step function of the number of batteries recharging at the same time. The time required for each battery to be guaranteed operational for $\tau_\text{fly}$ must satisfy $
   \max{(E_{j,h,i}(k_h,D_\text{max}))} \leq C_\text{bat} \,\,\, \forall \, j,h,i
$, where $C_\text{bat}$ should be kept to a minimum with some margin for errors. Therefore, the number of UAV batteries per single UAV that are required by the system is defined by the ratio $\tau_\text{charge}/\tau_\text{fly}$. The maximum number of replaceable on-board batteries is:
\begin{equation}
    b_\text{max}= \left\lceil n_\text{UAV} \left(\frac{\tau_\text{charge}}{\tau_\text{fly}}+1\right)\right\rceil,
\end{equation}
which has to be reflected in the purchasing price per UAV in the fleet.

\subsubsection{PV}
The solar energy generation units are represented by a set of photovoltaic (PV) panels placed in parallel, all of the same type \cite{PVdatasheet} and with the same working conditions. All parameters used in these equations are summarized in Table~\ref{PVparameters}. Their behaviour is simulated using a simplified version of the 5 parameters model \cite{5parammodel}, which neglects the shunt resistance and allows to calculate the maximum power voltage ($V_\text{m}$) and current ($I_\text{m}$) provided at any irradiation ($G^\text{irr}$) and ambient temperature ($T_\text{a}$) conditions. This is made possible using the conservative assumption of a maximum power point tracker (MPPT) with average efficiency $\epsilon_{\text{MPPT}}=95\%$ \cite{etaMPPT}.
\begin{IEEEeqnarray}{lCr}
\label{vmp}
V_{\text{m},h,i}=V_{\text{m},\text{ST}}-\beta\left(T_{\text{C},h,i}-T_{\text{ST}}\right)+V_{\text{t},h,i}\log{\frac{G^\text{irr}_{h,i}}{G^\text{irr}_{\text{ST}}}},\IEEEeqnarraynumspace\\
\label{vt}
V_{\text{t},h,i}= n_{\text{cells}}\frac{k_\text{b} n_\text{I} T_{\text{C},h,i}}{q},\\
\label{imp}
I_{\text{m},h,i}=I_{\text{m},\text{ST}}\left(\frac{G^\text{irr}_{\text{ST}}}{G^\text{irr}_{h,i}}\right)+\alpha \left(T_{\text{C},h,i}-T_{C,\text{ST}}\right),\\
\label{tcell}
T_{\text{C},h,i}=T_{\text{a},h,i}+\frac{T_{\text{C,NOC}} - T_{\text{a},\text{NOC}}}{G^\text{irr}_{\text{NOC}}}G^\text{irr}_{h,i}.
\end{IEEEeqnarray}
Knowing $V_\text{m}$ and $I_\text{m}$ from \eqref{vmp} and \eqref{imp}, as well as the cell temperature $T_\text{C}$, allows to calculate the output power as:
\begin{IEEEeqnarray}{lCr}
\label{PPV}
P_{\text{PV},h,i}(n_{\text{PV}})=n_{\text{PV}}\cdot V_{\text{m},h,i}\cdot I_{\text{m},h,i}\cdot \epsilon_{\text{conv}}\cdot\epsilon_{\text{MPPT}}.\IEEEeqnarraynumspace
\end{IEEEeqnarray}
In the above equations, the subscript ST means standard test conditions ($G^\text{irr}_{\text{ST}}=1000 W/m\textsuperscript{2}$, $T_{\text{a},\text{ST}}=25^{\circ}\text{C}$), whereas NOC stands for nominal operating conditions ($G^\text{irr}_{\text{NOC}}=800$ W/m\textsuperscript{2}, $T_{\text{a},\text{NOC}}=20^{\circ}\text{C}$). The cell temperature at standard test conditions $T_{\text{ST}}$, was calculated using \eqref{tcell}, but using $T_{\text{a},\text{SC}}$ and $G^\text{irr}_{\text{ST}}$ instead of $T_\text{a}$ and $G$. The list price for a single panel, pre-VAT, is \eur202 \, resulting in a PV system cost that scales linearly with the number of solar panels $n_\text{PV}$, as $F_\text{PV} = 202 \cdot n_\text{PV}$.

\begin{table}[]
\caption{PV Parameters from \cite{PVdatasheet}.}
\centering
\begin{tabular}{lll}
\hline
Label & Definition & Value \\ \hline
\hline
$\alpha$ & Thermal coefficient of SC current & 0.0474 \%/\textdegree C \\
$\beta$ & Thermal coefficient of OC voltage & -0.285 \%/\textdegree C \\
$\epsilon_{\text{conv}}$ & Converter efficiency & 95\% \\
$n_{\text{cells}}$ & Number of PV cells & 60 \\
$n_\text{I}$ & Diode ideality factor &  1.5\\
$k_\text{b}$ &  Boltzmann constant & $1.380649\cdot10^{-23}$ J/K\\
$q$ & Electrical charge of an electron & $1.602176634\cdot10^{-19}$ C \\
$G^\text{irr}_{\text{NOC}}$ & Irradiation at NOC & 800 W/m\textsuperscript{2}  \\
$G^\text{irr}_{\text{ST}}$ & Irradiation at ST & 1000 W/m\textsuperscript{2} \\
$I_{\text{m},\text{ST}}$ & Maximum power current at ST & 8.85 A  \\
$T_{\text{a},\text{NOC}}$ & Ambient temperature at NOC & 20\textdegree C \\
$T_{\text{C,NOC}}$ & Cell temperature at NOC & 45\textdegree C \\
$T_{\text{C,ST}}$ & Ambient temperature at ST & 25\textdegree C\\
$V_{\text{m},\text{ST}}$ & Maximum power voltage at ST & 31.8 V \\ 
$\epsilon_\text{MPPT}$ & Maximum Power Point Tracker & 95\%\\  \hline
\end{tabular}
\label{PVparameters}
\end{table}

\subsubsection{Wind}
\begin{figure}[t]
\setlength{\belowcaptionskip}{-5pt}
\centering
\centerline{\includegraphics[width=1.0\columnwidth]{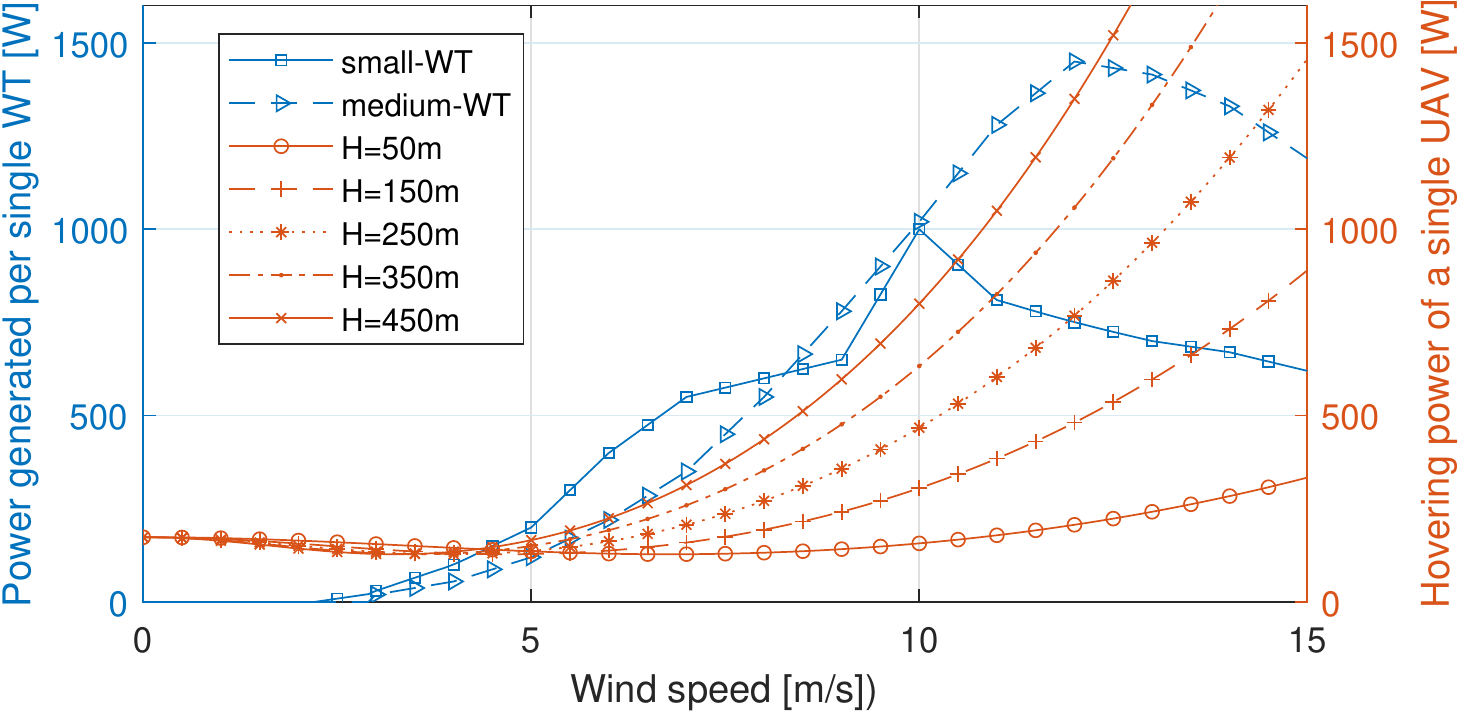}}
\caption{The wind power curves for two types of horizontal-axis WTs (blue), and UAV power consumption for hovering at different altitudes (red).}
\label{windcurve}
\end{figure}

In favor of precision, the power output of a wind turbine is not calculated with an analytical model, but by interpolating the generation data found in the data sheet \cite{WTdata}, and shown in blue in Fig.~\ref{windcurve}.
Two types of wind turbines are considered and treated as distinct elements of the system:
\begin{itemize}
    \item A horizontal axis small-WT with standard power output of 500W (unit cost $F_\text{W500} = $ \eur$1,429.95$);
    \item A horizontal axis medium-WT with standard power output of 1kW (unit cost $F_\text{W1000} = $ \eur$2,738.76$);
\end{itemize}
The list prices displayed above are pre-VAT, and were provided by Aeolos Wind Energy Ltd \cite{WTdata}. Such costs also include a 9\,m pole and a rectifier and control system. Finally, the total power output of the system is scaled to:
\begin{equation}
P_{\text{WT},h,i}(n_\text{W500},n_\text{W1000}) = \sum_a n_{a}P_{a}(v_{w,h,i}),
\end{equation}
where $a \in \{\text{W500},\text{W1000}\}$ depicts the type of the turbine out of the two suggested ones, and $n_a$ is the number of turbines of each type, giving a total cost of $F_\text{WT} = \sum_a F_{a} \cdot n_{a}$.
\subsubsection{Ground Battery}
In order to provide continuous service, an energy storage system with capacity $E_\text{cap}$ is needed. The load is powered by the generation elements when possible, with the storage elements receiving any excess energy and providing back-up when the power generated is too low. Therefore, at the end of a time slot the net energy in the system is $
    E_{\text{net},h,i} = \left[P_{\text{PV},h,i}(n_\text{PV})+P_{\text{WT},h,i}(n_\text{W500},n_\text{W1000})\right]\delta_t
    -E_{\text{UAVs},h,i}(k_h,D_\text{max}),$ 
where $\delta_t = 2 \tau_\text{fly}$ is the length of the time interval.
Thus, in case of net positive or negative energy, the battery capacity at the next time step $E_{(h+1)\%24,i+(h+1)/24}$ (where \% is the modulo operator and / is integer division) will increase or decrease by $\min(E_\text{cap}(n_\text{cell}),E_{h,i}+\epsilon_{\text{b},h,i} E_{\text{net},h,i}),$ with,
\begin{IEEEeqnarray}{lCr}
\label{etabat}
\epsilon_{\text{b},h,i}=\left\{\begin{matrix} \epsilon_\text{conv}, & \,\,\,  E_{\text{net},h,i} \geq 0\\
\frac{1}{\epsilon_\text{conv}}, & \,\,\,  E_\text{\text{net},h,i} <0
\end{matrix}\right.
\end{IEEEeqnarray}
where $E_\text{cap}$ is the total capacity of the battery as a function of the number of cells in the system $n_\text{cell}$, and $\epsilon_{\text{b},h,i}$ is the overall efficiency of the storage system.
The Li-ion battery cells are cylindrical LG MJ1, with unit cost of \eur5.75 and capacity of 12.6 Wh, resulting in a maximum ground battery (GB) capacity of $E_\text{cap}=12.6 \cdot n_\text{cell}$ requires spending $F_\text{E} = 5.75 \cdot n_\text{cell}$.

\section{Problem Definition and Methodology}
\label{problem}
The goal of this paper is to find the best system configuration for providing as much coverage as possible in a geographical region. We thus consider the problem of sizing the entire system as a combination of six variables: 1) number of communication UAVs in the fleet $n_\text{UAV}$; 2) number of 500W WTs $n_\text{W500}$; 3) number of 1kW WTs $n_\text{W1000}$; 4) number of PV panels $n_\text{PV}$; 5) number of battery cells in the GB $n_\text{cell}$; and 6) the radius of the circular CA $D_\text{max}$. In order to evaluate the quality of the system, we use the area $\pi D_\text{max}^2$, in which the guaranteed communications rate is satisfied, and the total upfront cost $F$ for the system needed to provide that service is lower than a given threshold.

{ 
Since accurately calculating the total capital expenditure of the system is crucial, the cost of the UAV swarm plays a big role. Since we use the DJI matrice 100/200 models as a reference, we take a reference price of \eur4000 per UAV, resulting in a total cost for UAV equipment of $F_\text{UAV} = 4000 \cdot n_\text{UAV}$. Moreover, this budget also covers spare batteries $b_\text{max}=3 \cdot n_\text{UAV}$ that are required for battery swapping. Finally, to guarantee operability in case of defects in one of the UAVs in the fleet, and to offer better interleaving for battery swapping \cite{nithinTVT}, there needs to be one spare UAV in the fleet $n_\text{UAV}\geq 2$. To avoid inconsistencies in the service, a simple timing difference in the UAV swapping time can be employed. To elaborate, not all UAVs have to do the battery swap at the same exact instant, as this would result in a total outage of the system. To circumvent this issue, the UAVs deployment is desynchronized by a few minutes. Finally, in case of a scheduling failure, the redundant UAV can substitute the designated UAV in the air.}

\begin{IEEEeqnarray}{rCl}
\text{(P1)} & : & \underset{\{n_\text{PV},n_\text{W500},n_\text{W1000},n_\text{cell},n_\text{UAV},D_\text{max}\}}{\text{maximize}}\,\,\,\, \frac{\pi D_\text{max}^2}{F}, \nonumber\\
\text{s.t.}& &  R_{\text{min},h}(D_\text{max}) \leq k_h R(k_h,D_\text{max}),  
\label{eq:ratebound}\\
&& 11 \geq n_\text{UAV} \geq 2,
\label{eq:numdrones}\\
&& n_\text{UAV} \geq \max_h{(k_h)}+1,
\label{eq:defK}\\
&& E_{h,i} \geq 0 ,
\label{eq:batterylim}\\
&& F = F_\text{PV}+F_\text{WT} +F_\text{UAV} + F_\text{E}\leq F_\text{max},
\label{eq:budget}\\
&& D_\text{max}\geq D_\text{lb},
\label{eq:lowbound}\\
&& D_\text{max} \leq D_\text{ub}.
\label{eq:upbound}
\end{IEEEeqnarray}
The (P1) objective function maximizes the coverage of the deployment normalized by its CAPEX; boundary \eqref{eq:ratebound} guarantees the quality of service for the whole area; boundary \eqref{eq:numdrones} maintains eligibility of the number of UAVs in the swarm; boundary \eqref{eq:defK} defines the size of the swarm; boundary \eqref{eq:batterylim} guarantees no system outage due to lack of energy; boundary \eqref{eq:budget} defines the financial budget; boundaries \eqref{eq:lowbound} and \eqref{eq:upbound} define the minimum and maximum required coverage. We note that if the problem is infeasible, the system is inadequate for the application scenario. Finally, as per \eqref{eq:ratebound} and \eqref{eq:batterylim}, the system does not allow for any outage tolerance given the provided data. {This is because mismanagement of energy allocation will not result in a total outage, but in a reduced quality of service. Since our goal is to provide average service to most users, accounting for outages would not be aimed towards constraining outages for that hour altogether, but a separate problem of QoS maximization instead of coverage maximization. So, in cases of sub-average performance of the system, it will operate in a best-effort mode.}

Given a fixed coverage area, the problem can be separated into two sub-problems that construct the CAPEX-efficient coverage maximization, and thus solution-searching can be done iteratively. The easier problem of the two is searching for the minimum energy load (MEL). 
\begin{IEEEeqnarray}{rCl}
\text{(MEL)} & : & \underset{\{k_h\}}{\text{minimize}}\,\,\,\, \min_{k_h} E_{ \text{UAVs},h,i}(k_h,D_\text{max}) \,\, \forall \, h,i, \nonumber\\
\text{s.t.}& & R(k_h,D_\text{max}) \leq R_{\text{min},h}(D_\text{max}), \\
&& 10 \geq k_h \geq 1.
\end{IEEEeqnarray}
The MEL problem guarantees coverage for a specific area by satisfying the lower datarate bound. Extracting the load profile of the entire system for a single area size is a constant complexity operation, since the number of hours and days for coverage are fixed. As such, the search of the entire space of eligible coverage areas has linear complexity, where the complexity of that operation scales with the size of the eligible space between both boundaries \eqref{eq:ratebound} and \eqref{eq:numdrones}. 
\begin{figure}[t]
\setlength{\belowcaptionskip}{-5pt}
\centering
\centerline{\includegraphics[width=\columnwidth]{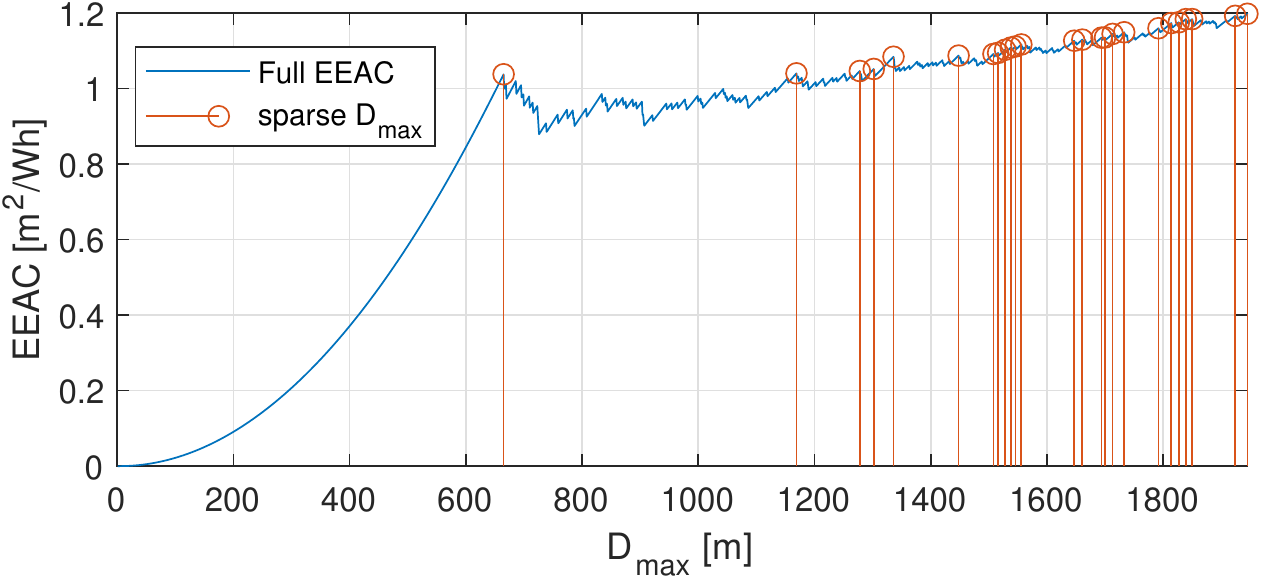}}
\caption{EEAC for servicing coverage area, at a suburban/remote setting, in the presence of weak wind with expected velocity of 3.6 m/s, overlay.}
\label{fig:yearlyenergyeff}
\end{figure}

The second sub-problem is finding the cheapest combination of energy elements (CCEE) W500 and W1000 WTs, PVs, and battery cells that satisfies the load profile. An exhaustive search on the CCEE problem has quartic complexity, which, summed with an exhaustively searched MEL, creates an unreasonably complex problem. In addition, checking the MEL + CCEE sub-problems for every possible coverage multiplies the complexity by the size of the space, $D_\text{lb} \leq D_\text{max} \leq D_\text{ub}$. Therefore, it is necessary to find a more efficient way to solve (P1). We approach this by performing greedy sparse search to reduce the solution space, and simpler algorithms to find near-optimal solutions. Approximate methods, such as the Genetic Algorithm (GA) implemented in \cite{marcogenetic}, did not yield a satisfying performance and are thus left out of this work. However, the computational performance of a GA is given at the end of Section~\ref{numerical}. 

{
To elaborate better, there are many challenges that come from solving the realistic design for a combination of a UAV swarm supplied with unreliable energy, such as renewables, in particular wind. The first challenge is that the swarm has a varying size during the day. The second challenge is that the energy expenditure has a non-monotone relationship with wind speed. The third challenge is that the energy generation also varies in a non-monotone manner with wind speed. However, the biggest challenge of all is that all aforementioned challenges do not scale linearly with the size of the coverage area. Thus, solving the optimal combination problem for all possible sizes of cellular coverage is non-tractable. Nonetheless, we developed an efficient way to find the most economical system configuration based on the service demand and the available resources whose data is region-specific and is obtained from European Commission's information system. The methods are elaborated in the following subsection.
}

\subsection{Greedy and Sparse Search (GSS) Algorithm}

\begin{algorithm}[t]
    \label{alg:SS}
\caption{GSS}
\textbf{IMPORT: \{MEL, BINARY-SEARCH, SAMPLE-mono, SAMPLE-2$^{nd}$der\,, SAMPLE-comb\}}\\
\textbf{Input}: all-constants, all-data, $D_\text{lb}$, $D_\text{ub}$ max\_budget; \\
j=0\\
$D_\text{max}=D_\text{lb}$\\
step\_size=1;\\
\While{$ D_\text{max} \leq D_\text{ub}$ \&\& $k_{h} \leq n_\text{UAV}$}
{j=j+1\\
(load$_{h,i}$\,[j] ,$F_\text{UAV}$\,[j]) \\=\textbf{MEL}$\left(E_{ \text{UAVs},h,i}(k_h,D_\text{max}) \,\, \forall \, h,i\right)$\\
EEAC\,[j]$= \pi D_\text{max}^2/\sum_{h,i} \text{load}_{h,i}\,\text{[j]}$\\
$D_\text{max}$ = $D_\text{max}$ + step\_size \\
}
\textbf{SAMPLE-mono}\,(EEAC,EEAC*):\\
EEAC$_\text{mnt}$ $\leftarrow$ \text{monotonic samples}\\ 
\textbf{SAMPLE-2$^{nd}$der}\, (EEAC$_\text{mnt}$,EEAC*):\\
$D_\text{max\_sparse},\text{addr} \leftarrow \text{ positive } 2^\text{nd} \text{ derivatives}$\\
j=0\\
\For{ $D_\text{max}$ $\mathbf{in}$ $D_\text{max\_sparse}$
}
{j=j+1\\
load$_{h,i}$ = load$_{h,i}$\,[addr[j]]\\
$F_\text{UAV}= F_\text{UAV}$\,[addr[j]]\\
$F_\text{comb} = 0$\\
$F=$\,max\_budget\\
solutions = []\\
flag==True\\
\While{flag==True}
{($n_\text{PV},n_\text{W500},n_\text{W1000}$),\,flag $\leftarrow$ \textbf{SAMPLE-comb} \\
$F_\text{comb} = F_\text{PV} + F_\text{WT}+F_\text{UAV}$\\
$n_\text{cell} = \lfloor(F-F_\text{comb})/5.75\rfloor$\\
\If{$E_{h,i}(n_\text{cell},n_\text{PV},n_\text{W500},n_\text{W1000},\text{load}) \geq 0 $
}
{\textbf{BINARY-SEARCH}$_{\text{minimize}} \,\, n_\text{cell}$ \, \textbf{s.t.}\\
$\,\,\,\,\,E_{h,i}(n_\text{cell},n_\text{PV},n_\text{W500},n_\text{W1000},\text{load})\geq 0$\\
$F_\text{comb} = F_\text{PV} + F_\text{WT}+F_\text{UAV}+F_\text{E}$\\
\textbf{APPEND} 
\,\,\, ($D_\text{max},n_\text{cell},n_\text{PV},n_\text{W500},n_\text{W1000},n_\text{UAV}, F_\text{comb}$) \\
    \textbf{TO} solutions\\
$F = F_\text{comb}$\\
}
}
fin\_sols[j] = $\min_{F_\text{comb}}$(solutions)
}
\textbf{Output}: fin\_sols
\end{algorithm}

We define a search algorithm that uses sparse searching of coverage areas where maximum coverage per unit cost is likely to occur, and uncover a simplified way to solve CCEE. Specifically, we investigate the energy efficiency of annual coverage (EEAC) for each size of coverage area as a proxy-heuristic metric:
\begin{equation}
    \text{EEAC} = \frac{\pi D_\text{max}^2}{\sum_{h,i} E_{\text{UAVS},h,i}},
\end{equation}
where $ E_{\text{UAVS},h,i}$ is given by the MEL problem. Looking at  Fig.~\ref{fig:yearlyenergyeff}, it is noticeable that EEAC is neither a monotonic nor a convex function of the coverage area. Therefore, it is convenient to sparsely search for a solution where EEAC is improving. Furthermore, we can use a greedy approach to shrink the number of samples that will be searched for a solution to the ones that offer the best improvement with regards to the last sample. Thus we select only the samples whose second order derivative is larger than zero. In this way, we still solve the MEL problem for the whole $D_\text{lb} \leq D_\text{max} \leq D_\text{ub}$ space beforehand, with the goal of reducing the search space for the multi-variate CCEE sub-problem by a significant factor ranging between 100-1000, depending on the scenario we investigate. 

{The CCEE problem is harder to simplify. However, we can exploit the fact that the budget can be dedicated to two different purposes: energy generation and storage. We can easily reduce the complexity of searching the viability of storage once we have sufficient energy generation supporting the system. Thus, we decrease the dimensionality of the space by increasing the budget until battery storage becomes relevant, i.e. the power generation profile is able to keep up with the load profile. Since we are looking to minimize the financial cost, the first eligible solution where battery storage is relevant becomes our new and smaller search space. A simplified representation of the algorithm is shown in Alg.~\ref{alg:SS}. The GSS approach does not guarantee to always find the global maximum for the coverage area due to the sampling of $D_\text{max}$. Despite this, GSS managed to find the global optimum for all the scenarios that we tested. This is mostly due to the well sampled areas and the exhaustive search \textbf{SAMPLE-comb} function for sampling combinations of WTs and PVs, which are monotonously increasing in cost.}

\section{Numerical Results and Case Analysis}
\label{numerical}
\begin{figure}[t]
\setlength{\belowcaptionskip}{-5pt}
\centering
\centerline{\includegraphics[width=\columnwidth]{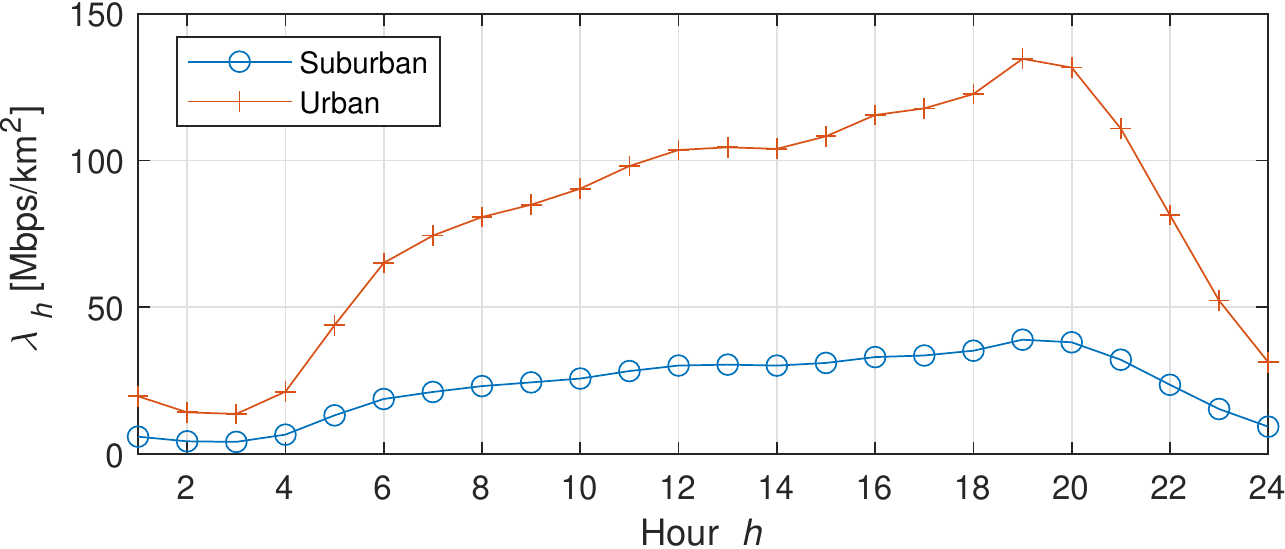}}
\caption{Daily evolution of requested data traffic \cite{marsan,donevski2019neural}.}
\label{fig:datarates}
\end{figure}
We aim to accurately evaluate the feasibility of the system across seasons or years. However, due to data sensitivity, the traffic data requested by the populace is only reflected on a daily cycle $\lambda_{h}$ and does not vary with location, as shown in Fig.~\ref{fig:datarates}. Moreover, we distinguish two possible types of areas that may need coverage: Suburban and Urban. These have the $(a,b,\eta^{\text{LoS}},\eta^{\text{NLoS}})$ propagational parameters of values $(4.88,0.43,0.2,24)$ and $(9.61,0.16,1.2,23)$, for Suburban and Urban respectively \cite{atg}. The  rest of the testing parameters are included in Table~\ref{tab:testparams}. Antenna directivity $A_\text{eff}$ is considered as a split variable, since it may impact the aerodynamics of the UAV in ways that the power consumption model cannot predict, and a system integrator may only have a few available types. 
\begin{figure}[t]
\setlength{\belowcaptionskip}{-5pt}
\centering
\centerline{\includegraphics[width=0.95\columnwidth]{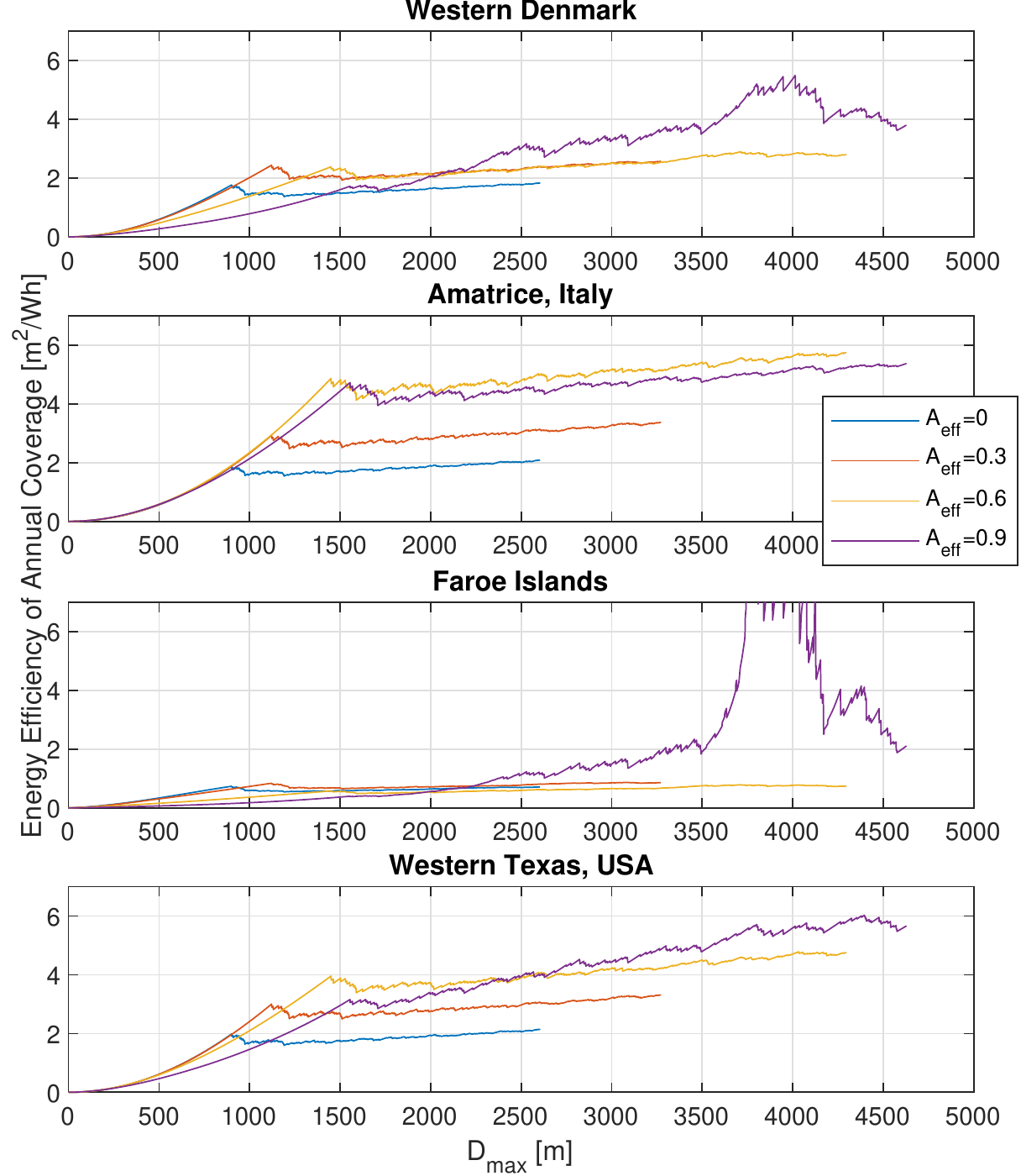}}
\caption{EEAC for all four locations in a suburban environment.}
\label{fig:allantennas}
\end{figure}
\begin{figure*}[t]
\setlength{\belowcaptionskip}{-5pt}
\centering
\centerline{\includegraphics[width=0.95\linewidth]{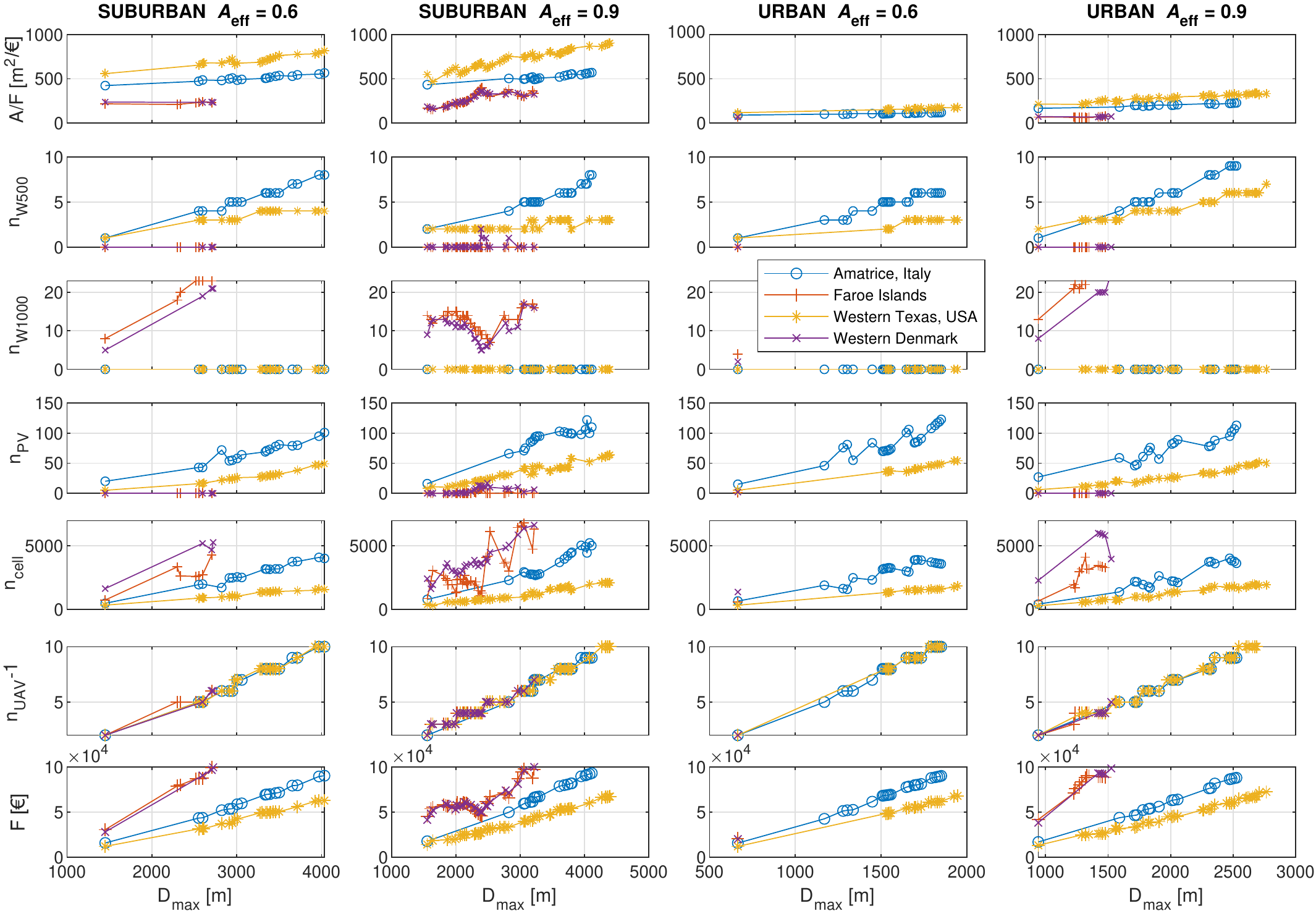}}
\caption{Implementation feasibility of the off-grid system in a suburban or urban environment.}
\label{fig:final1}
\end{figure*}

For the case analysis, four testing locations with diverse wind speed and solar irradiation patterns were chosen. Two locations are in regions that are prone to grid and system failures, like the earthquake ridden region around the Italian town of Amatrice and the fjord/floodplains of Western Denmark. We also suggest the placement of the system in common off-grid locations, such as sparsely populated areas in Western Texas and the touristic region of the Faroe Islands. We also refer to the Faroe Islands and Western Denmark as windy locations, and  Amatrice and Western Texas as sunny locations.

\begin{table}[]
\caption{Simulation Parameters \cite{URLLCdrone,9400376}}
\centering
\begin{tabular}{lll}
\hline
Label & Definition & Value \\ \hline
\hline
$f_\text{c}$ & Channel carrier frequency & 5.8 GHz \\
$c$ & Velocity of light & $3 \cdot 10^8$ m/s \\
$B$ & Channel bandwidth & 80 MHz \\
$H_0$ & Nominal height for wind measurements & 10 m \\
$E_{\text{w}}$ & Environment roughness coefficient & 0.335 \\
$B_\text{tot}$ & Available spectral width & 480 MHz \\
$N_0$ & Noise spectral power & -174 dBm/Hz \\
$P_\text{t}$ & Transmission Power & 23 dBm \\ 
$D_\text{lb}$ & Lower bound of coverage size & 0 m \\
$D_\text{ub}$ & Upper bound of coverage size & $\infty$ m \\
${P_\text{charge}}$ & Power of the charger & 180 W \\
$F_\text{max}$ & Total expendable budget & \eur100000  \\ \hline
\end{tabular}
\label{tab:testparams}
\end{table}

In Fig.~\ref{fig:allantennas} we illustrate the EEAC across the four scenarios for four different antenna directivity coefficients $A_\text{eff}$ that have negligible impact to the UAVs' aerodynamics. More efficient antennas expect higher optimal altitudes, thus consuming more energy for vertical flights and expecting higher wind velocities. Additionally, a larger $D_\text{max}$ implies larger swarms, that have a lower flying altitude. As discussed in the previous section, thanks to the aerodynamics of the UAV, the low speed of horizontal wind reduces the power consumption for hovering and proves beneficial for the flying swarm. This effect provides interesting results in the case of the more windy locations, such as the Faroe Islands, which shows a distinct improvement in energy efficiency for the coverage of areas with $3500\, \text{m} \leq D_{max} \leq 4250 \,\text{m}$.

{
All implementations are tested for full annual service on a specified location. This is the most difficult test for the system as it considers diverse weather patterns of all four seasons. The productivity of each energy source, wind harvested by WTs and solar harvested by PV panels, is tied to the geographical location of the CS and the time of year. Therefore, we use measurement-driven data provided by the European Commision's  Photovoltaic Geographical Information System\footnote{https://re.jrc.ec.europa.eu/pvg\_tools/en/\#MR} to extract the annual measurements of 2015. This way, by avoiding multi-year averages, correlation between hourly samples is retained. Moreover, we expect solar irradiation and wind speed to be inversely correlated, as per the study in \cite{widen2011correlations}. 

Since we have set the bound for CA as infinite, in Fig.~\ref{fig:final1} we plot the entire solution space of eligible area CA sizes searched by the GSS. This is done to better illustrate how the cost efficiency varies when different coverage bounds are imposed. The first impressions are that there is an obvious advantage in installing the system in an area where the use of solar panels is feasible. In both sunny locations, the cost-feasibility of the system is improved by better allocating the available budget, which in turn allows for exploiting the improved packing efficiency when using bigger swarms. The most cost efficient deployments are found in the Texan planes, where energy can be captured through both wind and solar technologies. Namely, bigger deployments in this setting do not need as much PV panels nor wind turbines as the other three locations to satisfy the energy requirements of the UAV swarm. 

Analyzing the impact of wind, we find that deployments in windy locations tend to have more volatile solution space, as opposed to the two sunny locations. The fluctuations in the curves as the CA gets larger are due to three non trivial effects of wind onto the energy generation and expenditure of the entire system. The first effect is the combination of the three non-monotonous curves for energy generation/expenditure in Fig.~\ref{windcurve}. The second effect that causes the strong fluctuations are the variability in wind speed for energy generation, which in some cases warrants use of solar panels to satisfy average QoS. The final effect is that denser deployments (when CA is larger it requires a bigger UAV swarm) tend to have a lower flying altitude. This is particularly impactful in windy environments, as the wind gets logarithmically stronger with the flying altitude, and thus increases the energy expenditure of the swarm. 

Furthermore, both windy locations do not find use for PV panels, and generally tend to use the bigger 1000W turbines. The first reason for this is that wind turbines offset the added wind expenditure of UAVs hovering in high winds; secondly, it makes more sense to exploit the natural resource with the higher energy output.  We note that, the behaviour between W500 and  W1000 WTs is in fact non-linear (it has a specific profile) and W1000 can operate very efficiently in higher winds, as it was shown back in Fig.~\ref{windcurve}. To add, the UAV energy expenditure initially dips for weaker winds, as it negates negative effects from poor aerodynamics. This means that W1000 turbines are better suited for cancelling the high energy expenditure of winds stronger than 10 m/s. Even in the cases of sunny deployments, a certain degree of wind power is necessary, usually with the W500 WTs. This is done to offset the higher energy expenditure in scenarios where the wind-speed becomes more challenging, such as when hovering at higher altitudes. As such, the flying altitude is the main culprit for the comparative efficiency between windy and sunny locations when the altitude of the UAVs becomes high, and wind speeds become the main cause of battery exhaustion. 

The analytic impact of service in urban and suburban situations varies due to two effects. The first is the difference in the propagation properties of the environment that impact the large scale fading. Tougher propagation environments, such as the urban environment, have more solid structures and thus expect higher hovering altitudes. The second effect is that urban-type environments expect larger data traffic requirements, as represented in Fig.~\ref{fig:datarates}, which results in smaller coverage areas for the same swarm size. }

As we can see in Fig.~\ref{fig:final1}, the difference in efficiency between the sunny and windy locations in suburban areas is not so drastic when the UAVs have antennas with high directivity $A_\text{eff}=0.9$. In accord, deployment efficiency between the windy and the sunny locations is much closer, mostly due to the higher altitudes of the UAVs. Nonetheless, UAV swarm deployments are more costly in windy locations and generally require much higher budgets than the \eur100000 to achieve coverage areas over 2700\,m or 3200\,m in radius when using $A_\text{eff}=0.6$ or $A_\text{eff}=0.9$ respectively. We note that this is an already high cost and thus we do not recommend the use of UAV swarm wireless communications for covering big windy areas. In contrast, the scenario in Western Texas is not limited by the upper CAPEX limit, due to the mild wind presence and reliable solar irradiation. 

Moving over to the Urban environment, we can see that, due to the increased user density and the worse propagation environment, the coverage for the urban implementations usually has much smaller CA of radius between $2000-2700$\,m. Moreover, the implementations' efficiencies of the windy and sunny locations are much closer. However, the budget of scaling the system to a bigger coverage area in a windy location takes a great toll on CAPEX and exceeds the initial allocation of $F=\text{\eur}100\,000$. Therefore, serving windy urban areas sees no use for the case of $A_\text{eff}=0.6$. Again, as seen in Fig.~\ref{fig:final1}, the performance for the windy location improves when using better antennas, allowing for larger swarms. 

{
It is important to note that the reliability of the system entirely depends on the data taken in the tested environments. In this work, we use four different sets of statistical data points, wind intensity, solar irradiation, temperature, data traffic pattern, based on averages over an hour. Therefore, the solutions provided by our analysis will give average estimates on communication service performance. This means that an unexpected fluctuation will not result in total system failure, but only in below-average performance in the communication service for that time slot. In the same way, unexpected peaks in the energy supply will make the system capable of providing better service than needed, and therefore provide above-average performance. These fluctuations in service are to be expected in an off-grid system, and due to the lack of available data and complex interactions we can address in the following way. 

\begin{figure}[t!]
    \centering
    \includegraphics[width=1\linewidth]{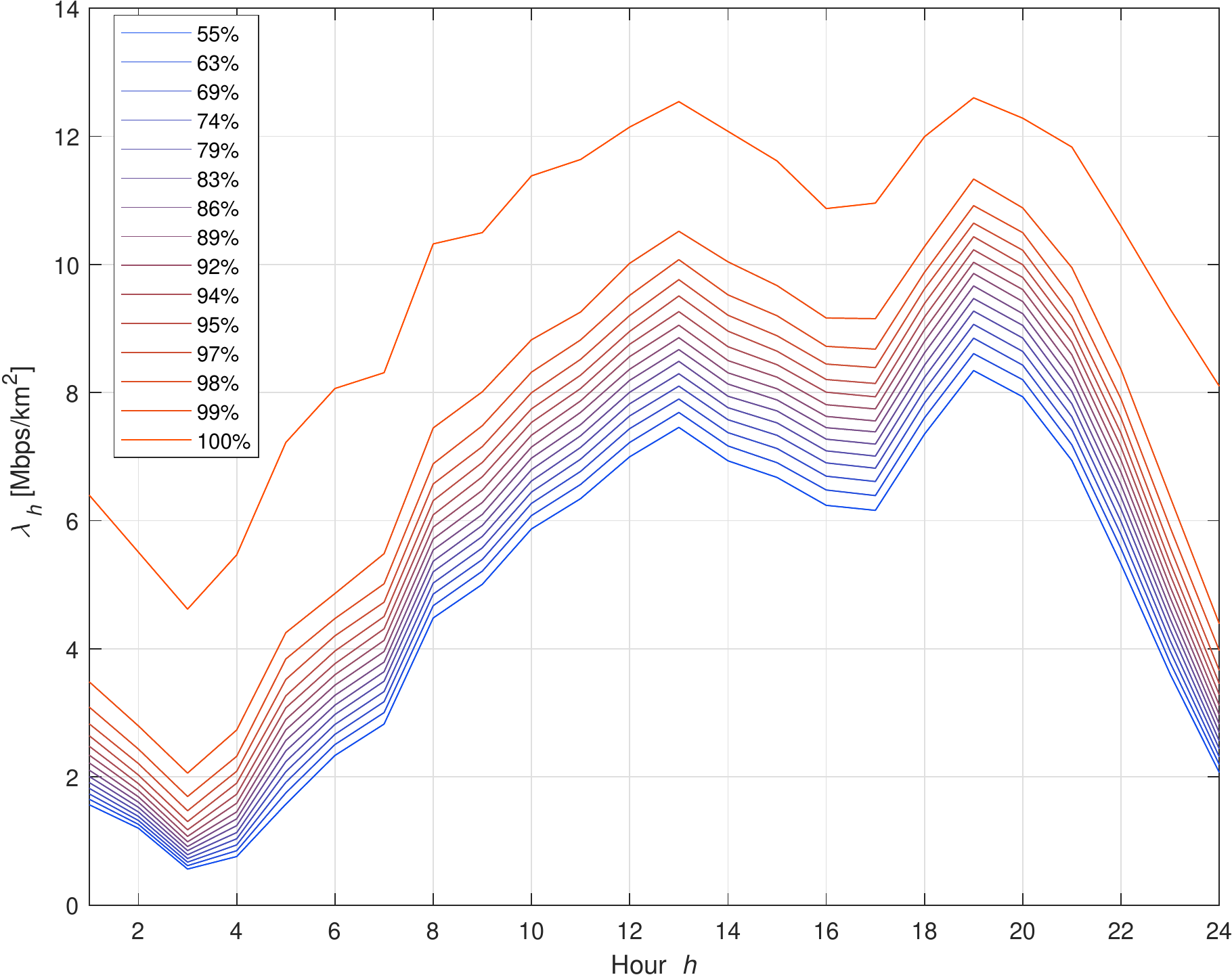}
    \caption{
    The expected service rate offering for 15 different levels of $\lambda_{h}$ provision ($x$ in P$(\lambda_{h}<x)$) for the suburban setting.
    }
    \label{fig:confidence}
\end{figure}
In the current results, the system design is targeted at offering normally provisioned service. In detail, given that $\lambda_{h}$ is a random variable with mean $\mu_{h}$ the aforementioned analysis covers the case where $\lambda_{h}=\mu_{h}$ Mbps/m$^2$. Regardless, the pre-deployment system analysis is designed to give more significance to better resource allocation through overprovisioning, or give more weight to the financial aspect by underprovisioning. As this work is focused on finding the optimal scale of deployment that makes the most financial sense for offering resource $R_{h,\text{min}}(D_\text{max}) = \lambda_{h} \pi D^2_\text{max}$, the uncertainty provisioning is embedded through sampling the empirical distribution of $\lambda_{h}$. To elaborate, improving the system for robustness to day-to-day variations in datarate and/or energy outage implies sampling the datarate request distribution at different points in the empirical cumulative distribution function P$(\lambda_{h}<x)$. In Fig.~\ref{fig:confidence} we showcase the above-average ($x>\mu_{h}$) resource demand sampled for 15 different levels of $x$ to illustrate the variability in resource provision.

Finally, we compared the performance of our custom made GSS algorithm to an exhaustive search solution and a GA implementation. An exhaustive search solution is the equivalent to independent planning where: first the UAV swarm configuration is solved for some specific area size, then the micro-grid system cost is minimized for that configuration, and the process is repeated for all possible area sizes. The performance comparison is in terms of processing time on the same machine with CPU execution, where the CPU was Intel(R) Xeon(R) Silver 4208 CPU @ 2.10GHz. The processing time for the case analysis of Western Texas was 51.88 minutes for GSS. An unrestrained exhaustive search solution containing the same solution space resolution as GSS was calculated to take approximately 3.64 years, which is unreasonable. A more practical, reduced resolution exhaustive search that was used to verify the GSS results took roughly 14 days of computation. A comparison with a GA with the following parameters was also carried out: population per generation of 30, number of parents in every generation 5, number of mutating offspring (in addition to the rest) 15, minimum consecutive generations when goal is reached 5, and minimum improvement desired by user 0.001 (the algorithm stops after no improvement higher than 0.001 has been made for 5 consecutive generations). The GA approach took 3.1 days of computation while providing results that were far from the optimal values.}

\section{Conclusion}
\label{conclusion}
In this paper we considered deploying a UAV swarm that offers persistent wireless services in an entirely off-grid setting. We formulated the problem as CAPEX efficient coverage area maximization, which is a multi-variate optimization problem for solving the load profile based on real world data. We considered energy generation from two sources, wind and solar, which are also taken from real world data. In this paper we have emphasized the importance of accounting for the impact of wind onto the deployment. Moreover, we consider the hourly wind intensity as a function of elevation and terrain roughness, and account for its impact on UAV deployments for long duration hovering. We have proposed the GSS algorithm, which is computationally easy, combining greedy sampling and binary search to find the optimal combination of wind turbines, PV panels, cells in the ground battery, and UAVs in the swarm. Using GSS we have calculated the feasibility of the system in four different locations where the deployment would have to balance wind or solar power generation. This work opens a plethora of directions for future works, such as investigating the feasibility in very specific areas, specific short term periods, and different types of UAVs. Additionally, a future direction can cover post-deployment energy management system for service rate matching.

\bibliographystyle{IEEEtran}
\bibliography{./bibliography.bib}

\end{document}